\documentstyle[aps,twocolumn,eqsecnum]{revtex}
\begin{document}
\title{{\small \noindent hep-th/9511098 \hfill USC-95/HEP-B5, TAUP 2303-95}
 \bigskip\\
U-Duality Multiplets \\
and Non-perturbative Superstring States}
\author{
Itzhak Bars\thanks{%
Research supported in part by the DOE Grant No. DE-FG03-84ER-40168. }\\
Department of Physics and Astronomy, \\
University of Southern California, \\
Los Angeles, CA 910089-0484, USA.\\
and \\
Shimon Yankielowicz\thanks{%
Research supported in part by the US-Israel Binational Science Foundation
and by GIF-the German-Israel Foundation for scientific research and the
Israel Academy of Science } \\
Department of Physics and Astronomy, \\
Tel-Aviv University, Ramat Aviv, Israel.
}
\maketitle

\begin{abstract}
\centerline {\bf ABSTRACT} \smallskip
{We employ an algebraic approach for unifying perturbative and
non-perturbative superstring states on an equal footing, in the form of
U-duality multiplets, at all excited string levels. In compactified type-IIA
supertring theory we present evidence that the multiplet is labelled by two
spaces, ``index'' space and ``base'' space, on which $U$ acts without mixing
them. Both spaces are non-perturbative extensions of similar spaces that
label perturbative T-duality multiplets. Base space consists of all the
central charges of the 11D SUSY algebra, while index space corresponds to
represetations of the maximal compact subgroup $K$ $\subset $ $U$. This
structure predicts the quantum numbers of the non-perturbative states. We
also discuss whether and how $U$-multiplets may coexist with 11-dimensional
multiplets, that are associated with an additional non-perturbative 11D
structure that seems to be lurking behind in the underlying theory.}
\end{abstract}


\section{Introduction}

It has been conjectured that, in addition to T-duality and S-duality, string
theory may possess an even larger duality, U-duality \cite{ht}, that
contains both T and S.
There is plenty of evidence for the existence of S duality and
circumstential evidence for U-duality
\cite{ht}-\cite{circumstancial2}. These ideas have provided a
degree of re-unification of certain string theories that {\it \`a priori}
seemed to be different. It now appears that there is a single underlying
theory with many moduli fields, and that the various conformal string
theories are different perturbative starting points from various corners of
moduli space. However, each perturbative expansion misses non-perturbative
aspects that may have already been described by another perturbative
expansion. Furthermore, beginning with 11D supergravity in the form of the
low energy limit, there seems to be an 11 dimensional structure lurking
behind the non-perturbative aspects of these string theories \cite{witten},%
\cite{townsend}-\cite{jhs2}. Further study of duality is bound to reveal
more non-perturbative properties of the underlying theory.

In this paper we investigate the multiplet structures for T and U-duality
transformations and use it as a device for unifying the perturbative and
non-perturbative states of the theory. This allows us to put all the states
on the same footing. Our discussion sheds new light into the
non-perturbative symmetry structure of the underlying theory and provides an
algebraic tool for discovering the non-perturbative states.

We will mainly discuss the case of the 10-dimensional type-IIA superstring
toroidally compactified on $R^d\times T^c,$ that has a $d$-dimensional
Minkowski spacetime with $c=10-d$ dimensions compactified on tori. The $T$%
-duality group is \cite{t-review}
\begin{equation}
T=O(c,c;Z)
\end{equation}
in all cases. The conjectured non-compact $U$-groups, their maximal compact
subgroups $K\subset U,$ and the maximal compact subgroup $k$ of the T-group,
\begin{equation}
k=O\left( c\right) \times O(c)
\end{equation}
are listed for various dimensions in the following table \cite{ht}. Note
that since $T\subset U$ then $k\subset K.$ It is understood that these
groups are continuous in supergravity but only their discrete version can
hold in string theory.
\begin{eqnarray*}
&&
\begin{tabular}{|lccc|}
\hline
\multicolumn{1}{|l|}{$d/c$} & \multicolumn{1}{c|}{$U$} & \multicolumn{1}{c|}{%
$K$} & $k$ \\ \hline
\multicolumn{1}{|l|}{9/1} & \multicolumn{1}{c|}{$SL(2)\otimes SO(1,1)$} &
\multicolumn{1}{c|}{$U(1)\otimes Z_2$} & $Z_2\otimes Z_2$ \\ \hline
\multicolumn{1}{|l|}{8/2} & \multicolumn{1}{c|}{$SL(3)\otimes SL(2)$} &
\multicolumn{1}{c|}{$SO(3)\otimes U(1)$} & $U(1)\otimes U(1)$ \\ \hline
\multicolumn{1}{|l|}{7/3} & \multicolumn{1}{c|}{$SL(5)$} &
\multicolumn{1}{c|}{$SO(5)$} & $SO(4)$ \\ \hline
\multicolumn{1}{|l|}{6/4} & \multicolumn{1}{c|}{$SO(5,5)$} &
\multicolumn{1}{c|}{$SO(5)\otimes SO(5)$} & $SO(4)\otimes SO(4)$ \\ \hline
\multicolumn{1}{|l|}{5/5} & \multicolumn{1}{c|}{$E_{6\left( 6\right) }$} &
\multicolumn{1}{c|}{$USp(8)$} & $Sp(4)\otimes Sp(4)$ \\ \hline
\multicolumn{1}{|l|}{4/6} & \multicolumn{1}{c|}{$E_{7\left( 7\right) }$} &
\multicolumn{1}{c|}{$SU(8)$} & $SU(4)\otimes SU(4)$ \\ \hline
\multicolumn{1}{|l|}{3/7} & \multicolumn{1}{c|}{$E_{8\left( 8\right) }$} &
\multicolumn{1}{c|}{$SO(16)$} & $SO(7)\otimes SO(7)$ \\ \hline
\end{tabular}
\\
&&\text{TABLE I. Duality groups and compact subgroups.}
\end{eqnarray*}

T-duality was originally understood in toroidal backgrounds, but the scope
of T-duality in string theory is much larger and it exists in more
complicated curved backgrounds involving both compact and non-compact
spacetimes \cite{t-review}, in particular in all gauged WZW models.
T-duality transformations act on perturbative string states, and it
can be verified that T-duality is valid order by order in string
perturbation theory as well as in string field theory \cite{kz}. Target
spaces related to each other by T-duality give the same physical results for
T-invariant quantities such as the partition function. Different string
target spaces are the analogs of different vaccua, and T-duality are the
analogs of large discrete gauge transformations that relate them.

The string states involved in the T-duality transformations are not all
degenerate in mass. Therefore, T-duality must be regarded as the analog of a
spontaneously broken symmetry, and the string states must come in complete
multiplets despite the broken nature of the symmetry.

In this paper we will first clarify the nature of T-multiplets of excited
string states and the properties of U-multiplets in low energy supergravity.
By putting well known results \cite{kz}\cite{jc} into a suitable form we
will emphasize the hints they provide for U-multiplets. Then we will use
this information as boundary conditions to investigate the structure of
non-perturbative U-multiplets, first in the supergravity sector ($l=0$),
including non-perturbative BPS saturates states, and then at all excited
string levels $l$. In particular we will explicitly construct the
U-multiplets for a number of low lying levels, up to level $l=5.$

To avoid later confusion we give a definition of level $l$ that applies to
both perturbative and non-perturbative states. The mass of perturbative
states includes contributions from Kaluza-Klein and winding numbers in
addition to the string excitation level $l.$ Thus, for perturbative states $%
l $ is the excitation level of {\it oscillators}, not the mass. For the
non-perturbative states we $define$ the level $l$ to be the same as the
level of the perturbative states to which they are connected by duality
transformations as explained below. Thus, $l$ is a label of the entire
U-multiplet. In the explicit structures that emerge, U-transformations, much
like T-transformations, do not mix perturbative or non-perturbative states
that belong to different levels (using the definition of level just given).
This is an empirical observation that holds in our investigation up to level
$l=5$, and thus we assume that it holds at all levels. Our analysis
critically depends on this assumption.

Our proposal is that a U-multiplet has the form
\begin{equation}
\Phi _{indices}^{(l)}(x^\mu ,base)  \label{fields}
\end{equation}
where the indices and the base form complete multiplets under $K$ and $U$
respectively, for every spin. In the rest of the paper we will collect the
evidence that leads to this structure and then use our proposal to make new
statements about the non-perturbative theory. Both the $index$ space and the
$base$ space will include non-perturbative extensions. We will show how the
full group $U$ acts on this structure without mixing the index and base
spaces at each level $l$.

Like in a T-multiplet, the states in a U-multiplet are not all
mass-degenerate. But unlike a T-multiplet, a U-multiplet can contain
perturbative states together with non-perturbative states ({\it minimal}
case), or only non-perturbative ones ({\it non-minimal} case). In the
minimal case, by knowing the structure of a U-multiplet as in (\ref{fields})
we can predict algebraically the quantum numbers of the non-perturbative
states by extending the quantum numbers of the known perturbative
T-multiplets that belong to the bigger U-multiplet. This completion provides
the minimal set of non-perturbative states necessary for U-duality. There
remains the non-minimal case, that includes the possibility of additional
purely non-perturbative, complete U-duality multiplets. With additional
input, our approach can provide information on some non-minimal
U-multiplets, as we will discuss in section-V.

Therefore, our approach is an algebraic method that should complement the
analytic methods of finding solutions for building up the non-perturbative
spectrum of the theory. The non-perturbative spectrum that we find
explicitly is one of the immediate outcomes of our approach.

In addition, our formulation hopefully provides hints for the
non-perturbative formulation of the theory. In particular, a challenging
question is whether U-multiplets can be made consistent with some deeper
underlying structure. A possible candidate is an underlying 11-dimensional
structure which has recently aroused much interest \cite{witten},\cite
{townsend}-\cite{jhs2}. We will find a certain amount of support for hidden
11-dimensions, but we will also raise some questions that require resolution.

\section{T-multiplets}

In this section we will show that the structure of perturbative T-multiplets
at all levels $l,$ is consistent with (\ref{fields}). In particular we want
to show that both $base$ and $index$ spaces transform and that they do not
mix with each other.

The bosonic string on $R^d\otimes T^c$ has the following {\it perturbative}
states at oscillator level $l$
\begin{equation}
\phi _{indices}^{\left( l\right) }\left( x^\mu ,base\right) \leftrightarrow
{\text{(oscillators)}}^{\left( l\right) }|p^\mu ,\vec{m},\vec{n}>
\label{tmult}
\end{equation}
where the $base\equiv \left( \vec{m},\vec{n}\right) $ are $c$-dimensional
vectors on the Lorentzian lattice $\Gamma ^{c,c}$ representing the
Kaluza-Klein and winding numbers, while the $indices$ are inherited from
oscillators. For the superstring (in the Green-Schwarz formalism) the
indices are also inherited from the Ramond vacuum, but without changing the
overall structure. As mentioned above, these states are clearly not all
degenerate in mass, however they are mixed with each other under T-duality
transformations. As we will see later we need to extend both the index and
base spaces to construct U-multiplets.

It is well known that $T=O(c,c;Z)$ acts linearly on the $2c$ dimensional
vector $\left( \vec m,\vec n\right) $  \cite{t-review}.
However it also acts on the indices
in definite representations. The action of $T$ on the indices is an induced
$k$-transformation that depends not only on {\it all} the parameters in $T$,
but also on the background $c\times c$ matrices ($G_{ij},B_{ij}$) that
define the tori $T^c$. To see this we analyze the T-transformations in more
detail. As is well known, $O(c,c;Z)$ acts linearly on the $2c$ dimensional
vector $\left( \vec m,\vec n\right) $ in such a way as to keep the following
dot product invariant
\begin{equation}
\left(
\begin{array}{cc}
\vec m_1 & \vec n_1
\end{array}
\right) \left(
\begin{array}{cc}
0 & 1 \\
1 & 0
\end{array}
\right) \left(
\begin{array}{c}
\vec m_2 \\
\vec n_2
\end{array}
\right) =\vec m_1\cdot \vec n_2+\vec n_1\cdot \vec m_2.
\end{equation}
$O(c,c;Z)$ transformations are characterized by $2c\times 2c$ integer
matrices of the form
\begin{equation}
M=\left(
\begin{array}{cc}
a & b \\
c & d
\end{array}
\right) ,\quad M^T\left(
\begin{array}{cc}
0 & 1 \\
1 & 0
\end{array}
\right) M=\left(
\begin{array}{cc}
0 & 1 \\
1 & 0
\end{array}
\right) .  \label{occ}
\end{equation}
Our interest here is in understanding the action of the $O(c,c;Z)$
transformations on the oscillators, and hence on the indices of the fields
in (\ref{fields},\ref{tmult}). This can be extracted from a study of $T$%
-invariance of string field theory \cite{kz}: $O(c,c;Z)$ acts in such a way
as to keep the left/right string excitation level numbers $N_L=\sum \alpha
_{-n}^{Li}G_{ij}\alpha _n^{Li}$ and $N_R=\sum \alpha _{-n}^{Ri}G_{ij}\alpha
_n^{Ri}$ invariant. It does so by transforming both the torus parameters ($%
G_{ij},B_{ij})$ and the oscillators as follows
\begin{eqnarray}
G^{\prime }+B^{\prime } &=&\left( a\left[ G+B\right] +b\right) \left(
c\left[ G+B\right] +d\right) ^{-1}  \nonumber \\
G^{\prime }+B^{\prime } &=&\left( -\left[ G+B\right] c^T+d^T\right)
^{-1}\left( \left[ G+B\right] a^T-b^T\right)  \nonumber \\
G^{\prime } &=&\left( V_L^{-1}\right) ^TGV_L^{-1}=\left( V_R^{-1}\right)
^TGV_R^{-1} \\
\alpha _n^{\prime Li} &=&\left( V_L\right) _j^i\,\alpha _n^{Lj},\quad \alpha
_n^{\prime Ri}=\left( V_R\right) _j^i\,\alpha _n^{Rj}  \nonumber \\
V_L &\equiv &\left[ c\left( G+B\right) +d\right] ,\quad V_R\equiv \left[
c\left( -G+B\right) +d\right]  \nonumber
\end{eqnarray}
where the second and first equations are equal by using (\ref{occ}). The
third equation is obtained from the first or second by projecting on the
symmetric part of both sides. So, the oscillators undergo transformations $%
V_L,V_R$ that depend on the $O(c,c;Z)$ parameters as well as on the target
space background fields $G,B.$

For our purposes we get a clearer picture by defining oscillators in a flat
basis rather than the ``curved'' basis of $G_{ij}.$ Thus, by introducing
vielbeins $G_{ij}=e_i^ae_j^a$ we have the flat oscillators
\begin{eqnarray}
\alpha _n^{La} &=&e_i^a\alpha _n^{Lj},\quad \alpha _n^{Ra}=e_i^a\alpha
_n^{Rj}  \nonumber \\
N_L &=&\sum \alpha _{-n}^{La}\alpha _n^{La},\quad N_R=\sum \alpha
_{-n}^{Ra}\alpha _n^{Ra}
\end{eqnarray}
whose transformation properties under $O(c,c;Z)$ amount to induced maximal
compact subgroup rotations$\,$ $O\left( c\right) _L\times O\left( c\right)
_R.$ This is seen by noting that orthogonal transformations are induced on
the flat index of the vielbein in order to maintain the transformation law
for the metric
\begin{eqnarray}
\left( e_i^a\right) ^{\prime } &=&\left( V_L^{-1}\right)
_i^j\,\,e_j^b\,\,\left( T_L\right) _b^a  \nonumber \\
\left( e_i^a\right) ^{\prime \prime } &=&\left( V_R^{-1}\right)
_i^j\,\,e_j^b\,\,\left( T_R\right) _b^a \\
T_L &\subset &O\left( c\right) _L,\quad T_R\subset O\left( c\right) _R
\nonumber
\end{eqnarray}
We emphasize that $T_L\neq T_R$ since $V_L\neq V_R,$ so that the left and
right moving flat oscillators $\alpha _n^{La},\alpha _n^{Ra}\,$ transform
under different orthogonal transformations.

Thus, for massive states the $indices$ must form complete representations
\begin{eqnarray}
indices\,\,\, &\longleftrightarrow &SO(d-1)\otimes k \\
k &=&O\left( c\right) _L\times O\left( c\right) _R.  \nonumber
\end{eqnarray}
where $SO(d-1)$ is the rotation group that identifies the spin of the state
in $d$-dimensions$.\,\,$In the $l=0$ sector the little group $SO(d-2)$
replaces the $SO(d-1)$ factor\footnote{%
Actually the type-II superstring massless states do form larger multiplets
under $SO(d-1).$ This is related to the fact that there is an 11D structure
in the type-II SUSY.}. We note that $k$ is larger than the naive $O(c)$ that
is embedded in $SO(10)\rightarrow SO(d-1)\otimes SO(c)$.

In the superstring theory of type-II, which is of interest in this paper,
the above discussion does not change in the presence of fermions. The
fermions are treated in the Green-Schwarz formalism. The ``vaccum'' state is
the Clifford vacuum for the zero mode fermions, that produces all the
massless states in compactified 11-dimensional supergravity. This set of
states, already form $U$-multiplets in supergravity by definition (see also
below), and therefore are consistent with the classification with $k\subset
K\subset U.$ Furthermore, the fermion oscillators end up in the spinor
representation of the maximal compact subgroup $k=O\left( c\right) _L\times
O\left( c\right) _R.$ Therefore, the states that they create at higher
levels $l$ are consistent with the index structure of $O\left( c\right)
_L\times O\left( c\right) _R.$ The upshot is that the superstring theory has
$T$-multiplets that are classified precisely as in (\ref{tmult}) for various
spins, including fermions. Later in section III, by explicit construction of
the states we will identify the $k$-multiplets in type II string theory at
each oscillator level.

\section{U-multiplets at level 0,\protect\\``index'' and ``base''}

In this section we study the structure of U-multiplets at level $l=0$
(including non-perturbative states) and show that they are of the proposed
form (\ref{fields}).

\subsection{Massless states and indices}

The low energy sector of the type-II\ string theory is described by
compactified 11-dimensional supergravity. The fields correspond to the
vacuum sector of the superstring in the Green-Schwarz formalism
\begin{equation}
\phi _{indices}^{(0)}(x^\mu )\longleftrightarrow |vac,p^\mu >
\end{equation}
The indices come from the R-R vacuum in the Green-Schwarz formalism. They
correspond to the short supermultiplet $2_B^7+2_F^7$ of 11D supersymmetry,
dimensionally reduced to $R^d\otimes T^{c+1}$. This is shown explicitly in
the appendix. Here one can start to re-classify the indices under $U$ by
using the original Julia-Cremmer classification \cite{jc} of the massless
states, as given in the appendix. It is seen that the fermions are not in $U$%
-multiplets, but rather in $K$-multiplets. Furthermore, the scalars
classified in the coset $U/K$ undergo a non-linear transformation under $U.$
So, the U-transformations that leave the supergravity Lagrangian invariant
act in non-linear ways.

The pattern of U-transformations is clearer in the description of
supergravity as a gauged sigma model. The scalars start out as a matrix $%
\exp \left( t\cdot \phi \right) $ in the adjoint representation of $U.$ One
must distinguishing {\it global} discrete transformation under the $%
U_{global}$-group that acts on one side of the matrix of scalars, and {\it %
local} maximal compact subgroup $K_{local}\subset U^{\prime }$
transformations that act on the other side of the matrix of scalars (where $%
U^{\prime }$ is isomorphic to $U$ but commutes with it). The fermions are
classified under $K_{local}$ while the massless vectors and massless
antisymmetric tensors are in representations of $U_{global}.$ The gauge
fields for the group $K$ are non-propagating, so one can choose a unitary
gauge and eliminate the auxiliary gauge fields through the equations of
motion. The remaining physical scalars are classified in the coset $U/K.$ In
the unitary gauge, the transformations of $U_{global}$ on the scalars have
to be compensated with transformations of $K_{local}$ in order to maintain
the gauge. The fermions transform under this field dependent $K_{local}$
whose free parameters are just the ones in $U_{local}.$ The massless vectors
and tensors continue to transform under $U_{global}.$ Thus, we outline the $%
l=0$ pattern of ``indices'' as follows
\begin{eqnarray*}
&&
\begin{tabular}{|l|l|l|}
\hline
$\text{spin}$ & $\left.
\begin{array}{c}
\text{action of } \\
U\text{-group}
\end{array}
\right. $ & $\text{representation}$ \\ \hline
$\text{scalars}$ & $U_{global}\otimes K_{local}$ & coset$\,\,\,U/K$ \\ \hline
$\text{vectors, tensors}$ & $U_{global}$ & dim. in$\text{ }$appendix \\
\hline
spinors & $K_{local}$ & dim.$\text{ }$in appendix \\ \hline
\end{tabular}
\\
&&\text{TABLE II. U-group acts in two ways in supergravity.}
\end{eqnarray*}

This classification is the starting point for U-multiplets. It already hints
that one must analyze the whole string spectrum from the point of view of
both global-$U$ and local-$K$ transformations that commute with each other.
However in the unitary gauge, which is the form in which string theory
presents itself, it is hard to keep track of the ``two sides'' of the matrix
of scalars. Therefore, in this paper we will be less ambitious and we will
explore some consequences of the {\it diagonal subgroup} $K_{diag}$ that
sits both in the global and local sides. It is this $K_{diag}$ subgroup that
classifies the ``indices'' of the fields in (\ref{fields}). We observe that
the classification of all the massless fields in table II is consistent with
(\ref{fields}) since all of them correspond to complete $K$-multiplets.

\subsection{Massive fields in $l=0$ multiplet}

Next consider all the massive perturbative string states of the form (\ref
{tmult}) in the $l=0$ sector that correspond to Kaluza-Klein and winding
solutions labelled by $\left( \vec{m},\vec{n}\right) $%
\begin{equation}
\phi _{indices}^{(0)}(x^\mu ,\vec{m},\vec{n})\longleftrightarrow |vac,p^\mu ,%
\vec{m},\vec{n}>.
\end{equation}
These have the same set of $indices$ as the massless fields corresponding to
the same $2_B^7+2_F^7$ short multiplet of 11D SUSY discussed above. These
massive fields can simply be regarded as if coming from the Kaluza-Klein
compactification of the supergravity fields, but extended by T-duality
transformations of the base. Evidently they exist in the string theory and
they form T-multiplets consistent with (\ref{fields}).

Now we discuss the non-perturbative BPS\ saturated soliton solutions which
have non-perturbative charges $z^I$. The existence of some such states is
proven by finding soliton solutions to the field equations of supergravity
\cite{ht}-\cite{circumstancial2}. This shows that the supersymmetry algebra
has non-trivial non-perturbative central charges $Z$ that commute both with
the momentum and supersymmetry generators.

The supercharges, momenta and central charges form an algebraic system which
must be quantum mechanically represented in the spectrum. That is, the
eigenstates of the commuting operators, including the central charges, {\it %
should} be included as labels of the states of the full theory. In our
discussion of the base space below we will show algebraically that the
non-perturbative $z^I$ are at the same footing as the perturbative $\left(
\vec m,\vec n\right) ,$ so the complete set of states in the $l=0$ sector
must be of the form
\begin{equation}
\phi _{indices}^{(0)}(x^\mu ,\vec m,\vec n,z^I)\longleftrightarrow
|vac,p^\mu ,\vec m,\vec n,z^I>
\end{equation}
where the indices again correspond to the $2_B^7+2_F^7$ short multiplet of
dimensionally reduced 11D SUSY. For fixed values of the $base$ this
multiplet correspond to BPS saturated states, hence the mass is a function
of the charges $\left( \vec m,\vec n,z^I\right) $. This provides an
algebraic construction of the states that were found in the form of
classical solutions. Actually, there are many more states, with definite
quantum numbers that are uniquely specified by U-transformations (given
below). Furthermore, it puts the perturbative and non-perturbative states on
the same footing. We will give below the precise U-transformation that mixes
them. Thus, the U-multiplet is in complete agreement with the form (\ref
{fields}).

A well known first example is the uncompactified theory with $\left(
d,c\right) =\left( 10,0\right) .$ There are no $\left( \vec{m},\vec{n}%
\right) ,$ but there is one non-perturbative central charge $z.$ The states
in the $l=0$ sector are $\phi _{indices}^{(0)}(x^\mu ,z)\longleftrightarrow
|vac,p^\mu ,z>.$ The classical solutions with non-trivial $z$ are known as
BPS saturated black holes \cite{ht}. Witten \cite{witten} interpreted this
set of fields as 11D supergravity fields compactified on $R^{10}\times S^1,$
thus recovering an additional dimension. The U-group in this case is
trivial, however this well known case serves to illustrate that our
reasoning so far is consistent with previous discussions. Furthermore, it
shows that there are signs of a hidden 11D structure at $l=0,$ and serves as
a starting point for investigating 11D structutre at higher levels $l$ \cite
{ib11d}.

\subsection{U-classification of the Base Space}

Let us now clarify the content and transformation properties of the $base$.
For the moment we concentrate on the BPS saturated black holes that are
0-brane solutions of the 11D supergravity equations, and return to the more
general $p$-branes at the end of the paper. In that case we label the base
as above
\begin{equation}
base=\left( \vec{m},\vec{n},z^I\right)
\end{equation}
We will show that these quantum numbers correpond to the central charges of
11D supergravity and at the same time that they are the sources that couple
to the massless vector fields of 11D supergravity listed in the appendix.
Combining the two statements we will learn the U-classification of the base.

Consider the supersymmetry algebra of type-IIA in 10D. There are two
supercharges of opposite chirality. Together they form the 32-component
spinor of 11D. In the compactified theory the 32 components are labelled as $%
Q_\alpha ^a$ where $\alpha $,$a$ label the spinor representations in the
Minkowski and compactified dimensions respectively. The centrally extended
superalgebra is
\begin{equation}
\left\{ Q_\alpha ^a,Q_\beta ^b\right\} =\delta ^{ab}\gamma _{\alpha \beta
}^\mu P_\mu +1_{\alpha \beta }Z^{ab}
\end{equation}
The $1_{\alpha \beta }$ denotes either a symmetric or antisymmetric
combination of Lorentz (spinor) indices into a Lorentz singlet. Then the
corresponding central charges $Z^{ab\text{ }}$are either symmetric or
antisymmetric respectively. Here we will show that the central charges $%
Z^{ab}$ form complete $K$ and $U$ multiplets and that these correspond
exactly to the same multiplets that describe the massless vector fields
listed in the appendix. This fact is proven by building Table-III in three
steps: (i) Classify the supercharges under $SO(d-1,1)\otimes k,$ noting that
$k=SO(c)_L\otimes SO(c)_R$ naturally emerges from the two chiral
supercharges, (ii) Construct all possible $Z^{ab}$ whenever a $1_{\alpha
\beta }$ is allowed by the Lorentz content, and note the $k$-classification,
(iii) Combine the $k$-representations precisely to complete $K$
representations, as listed.

\begin{eqnarray*}
&&
\begin{tabular}{|l|l|l|l|}
\hline
$d/c$ & $\left.
\begin{array}{c}
32\,\,\,\,Q_\alpha ^a,\text{ under} \\
SO(d-1,1)\otimes k
\end{array}
\right. $ & $\left.
\begin{array}{c}
Z^{ab} \\
\text{under }k
\end{array}
\right. $ & $\left.
\begin{array}{c}
Z^{ab} \\
\text{under }K
\end{array}
\right. $ \\ \hline
10/0 & $16_L+16_R$ & 1 & 1 \\ \hline
9/1 & $16_{+}+16_{-}$ & $\left( _{+}^{+},_{-}^{-},_{-}^{+}\right) $ & $%
\left( _{+}^{+},_{-}^{-},_{-}^{+}\right) $ \\ \hline
8/2 & $\left.
\begin{array}{c}
\left( 8_L^{+},\left( \pm \right) _L\right) \\
+\left( 8_R^{-},\left( \pm \right) _R\right)
\end{array}
\right. $ & $\left.
\begin{array}{c}
\left( _{+}^{+},_{-}^{-},_{-}^{+}\right) _L \\
\left( _{+}^{+},_{-}^{-},_{-}^{+}\right) _R
\end{array}
\right. $ & 3$_{+}+$3$_{-}$ \\ \hline
7/3 & $\left( 8,\left[ \left( 2,0\right) +\left( 0,2\right) \right] \right) $
& $\left.
\begin{array}{c}
\left( 3,0\right) \\
+\left( 0,3\right) \\
+\left( 2,2\right)
\end{array}
\right. $ & 10 \\ \hline
6/4 & $\left.
\begin{array}{c}
\left( 4_L,(4_{spin},0)\right) \\
+\left( \bar{4}_R,(0,4_{spin})\right)
\end{array}
\right. $ & ($4_{spin}$,$4_{spin}$) & (4,4) \\ \hline
5/5 & $\left( 4,\left[ \left( 4,0\right) +\left( 0,4\right) \right] \right) $
& $\left.
\begin{array}{c}
\left( 5,0\right) \\
+\left( 0,5\right) \\
+\left( 4,4\right) \\
+2\left( 0,0\right)
\end{array}
\right. $ & 27+1 \\ \hline
4/6 & ($4_{spin},[\left( 4,0\right) +(0,\bar{4})])$ & $\left.
\begin{array}{c}
\left( 6,0\right) \\
+\left( 0,\bar{6}\right) \\
+\left( 4,\bar{4}\right)
\end{array}
\right. $ & 28$_{%
\mathop{\rm complex}
}$ \\ \hline
3/7 & $\left( 2,\left[ \left( 8,0\right) +\left( 0,8\right) \right] \right) $
& $\left.
\begin{array}{c}
\left( 21,0\right) \\
+\left( 7,0\right) \\
+\left( 0,21\right) \\
+(0,7) \\
+\left( 8,8\right)
\end{array}
\right. $ & 120 \\ \hline
\end{tabular}
\\
&&\text{TABLE III. Classification of central charges}
\end{eqnarray*}

Now, by comparing to the U-representations of the vectors listed in the
appendix we see that both the counting and the $K$-representation content is
the same\footnote{%
In the case $\left( d,c\right) =\left( 5,5\right) $ we found one extra
central charge in addition to the 27, listed as 27+1 in Table-III. According
to the appendix, it seems that there is no corresponding singlet massless
vector in compactified 11D supergravity. This is curious.} as the $Z^{ab}.$
Hence, the central charges
\begin{equation}
Z^{ab}=\left( \vec{m},\vec{n},z^I\right)
\end{equation}
form a complete U-multiplet.

The central charges and the sources for the massless vectors are related as
follows. The Kaluza-Klein and winding charges $\left( \vec{m},\vec{n}\right)
$ are the ``perturbative charges'' among the $Z^{ab}$. In the NSR formalism
they couple to the massless vectors that come from the NS-NS sector. From
the point of view of 10D supergravity compactified on $R^d\otimes T^c$,
these massless vectors are the gravi-photon and its axial partner which come
from the dimensional reduction of the 10D metric $g_{\mu \nu }$ and
antisymmetric tensor $B_{\mu \nu }$
\begin{equation}
g_{\mu \nu }\rightarrow \vec{V}_\mu ,\,\,\quad B_{\mu \nu }\rightarrow \vec{V%
}_\mu ^{\prime },
\end{equation}
The non-perturbative 0-brane solutions have charges $z^I$ that couple to the
remaining massless vector fields $V_\mu ^I$. These vectors come from the
Ramond-Ramond sector in the NSR formalism. Thus, in string theory
compactified on $R^d\otimes T^c$ the base contains altogether $N_{d,c}$
charges for 0-branes, which is the same as the number $N_{d,c}$ of massless
vector fields in 11D supergravity compactified on $R^d\otimes T^{c+1}$.

The non-perturbative charges $z^I$ are now on the same footing as the
perturbative charges $\left( \vec{m},\vec{n}\right) $ since they both are
sources in the field equations of the massless vector fields. This suggests
naturally that the $base$ in (\ref{fields}) transforms in the same U-duality
multiplet as the vector fields themselves. This is precisely the multiplet
of size $N_{d,c}$ listed in the appendix. For example for $\left( d,c\right)
=\left( 6,4\right) $ the vectors fall into the 16-dimensional spinor
representation of $SO\left( 5,5\right) .$ Hence, the $base$ also transform
in the same linear representation\footnote{%
In the manifestly spacetime supersymmetric Green-Schwarz formalism of the
type-II string, all massless fields, including the massless vector fields,
come from the R-R vaccum sector that is described by only the fermionic zero
modes. In this sense the discussion so far has not involved any oscillators,
so that it is appropriate to use the term ``base'' for all the quantum
numbers included above.}.

\section{U-multiplets and string excitations at $l\geq 1$}

Our approach is the following: We start with $T$-multiplets that can be
defined perturbatively at every level $l.$ The $T$-multiplet indices form $k$%
-multiplets, which are in turn required to be part of complete larger $K$
multiplets in accordance with (\ref{fields}). This last part is the
consistency requirement for the $T$-multiplets to reassemble into $U$%
-multiplets. The base is already a U-multiplet as discussed above, so no
more discussion is needed. The remaining question is whether there is a need
to add non-perturbative $k$-multiplets (indices) to the perturbative ones in
order to find $K$ multiplets and thus satisfy $U$-duality. Here is a summary
of what we find, and which will be described below in several steps:

We will prove that at level $l=1$ the perturbative $k$-multiplets have
precisely the index structure that forms complete $K$-multiplets. No
additional indices are needed at level $l=1$, just like the case of level $%
l=0.$ This gives more credibility to our proposal in eq.(\ref{fields}). This
fact is quite non-trivial in various dimensions $\left( d,c\right) $, and we
will be able to explain it as a consequence of the underlying spacetime
supersymmetry.

At levels $l\geq 2\,\,$the requirement of complete $K$-multiplets predicts
the existence of additional non-perturbative $k$-multiplets beyond those
that can be created by applying oscillators on the ``base''. That is, more
``non-perturbative'' indices with predicted properties must be added. The
quantum numbers of these states are therefore completely determined. If
these additional states do not exist in the theory there is no $U$-duality$.$

\subsection{Perturbative k-multiplets in type II string at $l\geq 1$}

In order to obtain the perturbative $k$-multiplets explicitly we examine the
known spectrum of the type-IIA superstring in 10D before compactification,
and notice that there are some larger symmetry structures that help in
extracting directly the $k$ structure after compactification, as explained
below.

Our analysis begins by re-examining the perturbative spectrum and extracting
the indices
\begin{equation}
indices\longleftrightarrow (\text{Bose}\oplus \text{Fermi oscillators)}%
^{\left( l\right) }|vac>
\end{equation}
This was done up to level $l=5$ in \cite{ib11d}. The result for $d=10,$ $c=0$
is equivalent to a collection of fields $\phi _{indices}^{\left( l\right)
}\left( x^\mu \right) $ where the indices have the following structure of
representations
\begin{equation}
indices\Rightarrow \left( 2_B^{15}+2_F^{15}\right) \times R^{\left( l\right)
}  \label{10dpert}
\end{equation}
The factor $2_B^{15}+2_F^{15}$ represents the action of 32 supercharges on a
set of $SO(9)$ representations $R^{\left( l\right) }$ at oscillator level $%
l, $ where $SO(9)$ is the spin group in 10-dimensions for massive states.
The factor $R^{\left( l\right) }$ is of the form of direct products of $%
SO(9) $ representations coming from left/right movers
\begin{equation}
R^{\left( l\right) }=\left( \sum_ir_i^{(l)}\right) _L\times \left(
\sum_ir_i^{(l)}\right) _R
\end{equation}
such that the left-factor is identical to the right-factor, and is given by
the following collection of $SO(9)_{L,R}$ representations

\begin{eqnarray*}
&&
\begin{tabular}{|l|l|}
\hline
Level & SO(9)$_{L,R}$ representations $\left( \sum_ir_i^{(l)}\right) _{L,R}$
\\ \hline
$l=1\quad $ & $1_B$ \\ \hline
$l=2$ & $9_B$ \\ \hline
$l=3$ & $44_B+16_F$ \\ \hline
$l=4$ & $(9+36+156)_B+128_F$ \\ \hline
$l=5$ & $
\begin{array}{l}
\left( 1+36+44+84+231+450\right) _B \\
+\left[ 16+128+576\right] _F
\end{array}
$ \\ \hline
\end{tabular}
\\
&&\text{TABLE IV. L/R oscillator states of 10D superstring.}
\end{eqnarray*}
This structure shows that the factor $R^{\left( l\right) }$ is really
classified by the larger group
\begin{equation}
SO(9)_L\otimes SO(9)_R.
\end{equation}
Furthermore, the supercharge factor $2_B^{15}+2_F^{15}$ has an even larger
classification group
\begin{equation}
SO(32)
\end{equation}
with $2_B^{15}+2_F^{15}$ corresponding to the two spinor representations.
The diagonal $SO(9)$ subgroup of all these factors is the familiar rotation
group in the Lorentz group $SO(9,1)$.

When the the string theory is compactified to $R^d\otimes T^c,$ with $%
c+d=10, $ the set of indices in (\ref{10dpert}) needs to be re-classified by
\begin{equation}
SO\left( d-1\right) \otimes \left[ SO\left( c\right) _L\otimes SO\left(
c\right) _R\right]
\end{equation}
in order to identify the perturbative $k$-multiplets and their spins. Note
that this is a larger group than what is contained in the rotation group
alone $SO\left( 9\right) \supset SO\left( d-1\right) \otimes SO\left(
c\right) .$ Therefore, the larger symmetry structures that we identified
above are needed to obtain the $k$-multiplets. It is then evident that the $%
R^{\left( l\right) }$ factor, described by the representations in Table-IV,
is easily reduced in the form
\begin{equation}
SO(9)_{L,R}\Rightarrow SO\left( d-1\right) _{L,R}\otimes SO\left( c\right)
_{L,R}.
\end{equation}
For example, for $d=6,$ $c=4$%
\begin{eqnarray}
l &=&1:\quad 1\rightarrow 1  \nonumber \\
l &=&2:\quad 9_{L,R}\rightarrow \left( 5,1\right) _{L,R}+\left( 1,4\right)
_{L,R},\quad etc.
\end{eqnarray}
After this step the $SO\left( d-1\right) _L\times SO\left( d-1\right) _R$ is
reduced to the diagonal rotation group $SO(d-1)$ in order to give the final
classification of states. Naturally the outcome is the desired
classification under $SO\left( d-1\right) \otimes \left[ SO\left( c\right)
_L\otimes SO\left( c\right) _R\right] $.

Similarly, the $2_B^{15}+2_F^{15}$ supercharge factor may be reduced to the
representations of
\begin{equation}
SO\left( d-1\right) \otimes \left[ SO\left( c\right) _L\otimes SO\left(
c\right) _R\right]
\end{equation}
since we have already seen in Table-II that the 32 supercharges are already
classified under this group. Then the reclassification of the factor $%
2_B^{15}+2_F^{15}$ follows by taking the product of the supercharges.
Equivalently, specifying how to decompose the 32-dimensional vector of
SO(32), provides the instructions for decomposing the $2^{15}$ dimensional
representations as well.

Therefore all the perturbative indices in (\ref{10dpert}) do form $k$%
-multiplets that we can identify explicitly thanks to the larger structures $%
SO(32)$ and $SO(9)_L\otimes SO(9)_L.$ Combining this result with the base $%
\left( \vec{m},\vec{n}\right) $ proves that we have explicit control of the $%
T$-multiplets $\phi _{indices}^{(l)}(x^\mu ,\vec{m},\vec{n})$ up to level $%
l=5.$ We will next construct $U$-multiplets by starting with these $T$%
-multiplets.

\subsection{$U$-multiplets at $l=1$}

In the discussion above we used the explicit $SO\left( d-1\right) \otimes k$
classification of the 32 supercharges of Table-III. Here we go one step
further and give in Table-V the re-classification of the 32 supercharges
under the group $\left[ SO\left( d-1\right) \otimes K\right] $.

\begin{eqnarray*}
&&
\begin{tabular}{|l|l|}
\hline
$\frac dc\rightarrow \frac{d-1}{c+1}$ & $\left.
\begin{array}{c}
32\,\,\text{SUPERCHARGES}\,\,Q_\alpha ^a, \\
\text{under\thinspace \thinspace }SO(d-1)\otimes K
\end{array}
\right. $ \\ \hline
10/0$\rightarrow $9/1 & $16+16$ \\ \hline
9/1$\rightarrow $8/2 & $\left( 8_{+},\pm \right) +\left( 8_{-},\pm \right) $
\\ \hline
8/2$\rightarrow $7/3 & $\left( 8,2,\pm \right) $ \\ \hline
7/3$\rightarrow $6/4 & $\left( 4,4\right) +\left( 4^{*},4\right) $ \\ \hline
6/4$\rightarrow $5/5 & $\left( 4,4,0\right) +\left( 4,0,4\right) $ \\ \hline
5/5$\rightarrow $4/6 & $\left( \left( 2,0\right) ,8\right) +\left( \left(
0,2\right) ,8\right) $ \\ \hline
4/6$\rightarrow $3/7 & $\left( 2,8\right) +\left( 2,8^{*}\right) $ \\ \hline
3/7$\rightarrow $2/8 & $\left( \pm ,16\right) $ \\ \hline
\end{tabular}
\\
&&\text{TABLE V. Classification of 32 }Q_\alpha ^a\text{ exhibits }K
\end{eqnarray*}
Table-V was constructed in several steps: (i) The vector of $SO(32)$ is
identified with the 32-dimensional spinor of 11-dimensions. (ii) This spinor
is decomposed into the two 16-dimensional spinor representations of SO(10),
where SO(10) is the rotation group in 11-dimensions. (iii) The $SO(10)$
spinor representations are decomposed into the products of spinors of $%
SO(d-1)\times SO(c+1)$ where $SO(d-1)$ is the rotation group in $d$%
-dimensional Minkowski space, (iv) Then we find that the resulting spinor
representations of $SO(c+1)$ come together with the correct numbers to fit
into the complete $K$-representations given above.

The subgroup $K$ turns out to be the largest subgroup in $SO(32)$ that
remains after identifying the spin group $SO\left( d-1\right) $ as described
above. So, $K$ emerges in an interesting way as this largest factor. Each $K$%
-multiplet then contains several $k$-multiplets of the supercharges
automatically. For example, for $d=6,$ $c=4$ there are two supercharges
classified as $\left( 4,0\right) +\left( 0,4\right) $ under the $%
K=SO(5)\times SO(5)=Sp\left( 4\right) \times Sp\left( 4\right) .$ They could
be further re-classified under $k=SO(4)\times SO(4)$. Recall that our goal
is to reassemble all representations into $K$-multiplets. With this table we
have demonstrated that this is always the case for any representation of $%
SO(32)$ that is a product of the 32 supercharges. Hence the supercharge
factor $2_B^{15}+2_F^{15}$ admits a reclassification under $K$ for all
compactifications $\left( d,c\right) .$

There remains to discuss the $K$-reclassification of the $R^{\left( l\right)
}$ factor in the perturbative indices in (\ref{10dpert}) separately for
every level $l$. According to table III, at level $l=1$ we have just a
singlet
\[
l=1:\quad R^{\left( 1\right) }=1.
\]
Obviously, this is easily re-classified also as a singlet under $K.$
Therefore, for all compactifications $\left( d,c\right) $ we have now
demonstrated that level $l=1$ states do have indices that correspond to
complete $K$-multiplets. Hence at level $l=1$ we have $U$-multiplets of the
form (\ref{fields}) without needing any additional non-perturbative indices.
The only non-perturbative aspects at level $l=1$ come through the
non-perturbative charges $z^I$ at the base.

\subsection{U-multiplets at $l\geq 2$}

Next we analyze levels $l\geq 2.$ We know that the supercharge factor $%
2_B^{15}+2_F^{15}$ works, so we ignore it, and concentrate on the $R^{\left(
l\right) }$ factor. It is not straightforward to carry out the analysis
simultaneously for every $\left( d,c\right) .$ Therefore, we need to do it
one case at a time. It is very easy to analyze the case $\left( d,c\right)
=\left( 6,\,4\right) $ so we present it here as an illustration$.$ In this
case the spin group is $SO(5)$ and there are $4$ internal dimensions$.$ The
duality groups are
\begin{eqnarray}
U &=&SO(5,5),\quad K=SO\left( 5\right) \otimes SO\left( 5\right)  \nonumber
\\
T &=&SO(4,4),\quad k=SO\left( 4\right) _L\otimes SO\left( 4\right) _R
\nonumber \\
R^{\left( 2\right) } &=&\left( 5_{space}+4_{L,int}\right) \otimes \left(
5_{space}+4_{R,int}\right) , \\
\quad R^{\left( 3\right) } &=&etc.  \nonumber
\end{eqnarray}
where the indices $R^{\left( 2\right) }=9_L\otimes 9_R$ have been
reclassified according to their space and internal components. It is clear
from this form that the $k=SO\left( 4\right) _L\otimes SO\left( 4\right) _R$
structure follows directly from the left/right internal components. This
identifies the specific $k$-multiplets in $R^{\left( 2\right) }$ for various
spins. Can these $SO\left( 4\right) _L\otimes SO\left( 4\right) _R$
multiplets be reassembled into $SO\left( 5\right) \otimes SO\left( 5\right) $
multiplets for every spin? The answer is obviously no! This means that some
non-perturbative states are missing. The structure of the indices that are
missing corresponds precisely to increasing the
\begin{equation}
\left( 4_{int}\right) _{L,R}\rightarrow \left( 5_{int}\right) _{L,R}.
\label{4to5}
\end{equation}
This is equivalent to requiring non-perturbative states that correspond to
increasing the size of the $l=2$ multiplet in the form
\begin{equation}
R^{\left( 2\right) }=9_L\otimes 9_R\rightarrow 10_L\otimes 10_R.
\label{9to10}
\end{equation}
This result was found in \cite{ib11d} by assuming the presence of hidden
11-dimensional structure in the non-perturbative type-IIA superstring theory
in 10D. In ref.\cite{ib11d} a justification for (\ref{9to10}) could not be
given. However, in the present analysis $U$-duality demands (\ref{4to5}) and
therefore justifies (\ref{9to10}).

For $\left( d,c\right) =\left( 10,0\right) ,\left( 9,1\right) ,\left(
8,2\right) ,\left( 6,4\right) $ the analysis for $l=2,3,4,5$ produces
exactly the same conclusion as the 10D analysis, in that $U$-duality demands
that the $SO(9)_L\otimes SO(9)_R$ multiplets $R^{\left( l\right) }$ should
be completed to $SO(10)_L\otimes SO(10)_R$ multiplets. The minimal
completion is sufficient in this case. The necessary minimal completion was
given and discussed in \cite{ib11d}, where the possibility of additional
non-perturbative complete $SO(10)$ multiplets is also discussed. For all
higher levels $l=2,3,4,5$ there is a clear remarkable pattern of the {\it %
minimal} missing non-perturbative states. Their quantum numbers (for both
bosons and fermions) coincide systematically with the sum of all lower lying
perturbative states listed in Table-IV. This observation systematically
gives all the {\it minimal} states required by U-duality at all levels.
Hence, in these compactifications $U$-duality is consistent with a hidden
11D structure.

Next consider the example $\left( d,c\right) =\left( 7,3\right) .$ Now we
have a spin group $SO(6)=SU(4)$%
\begin{eqnarray}
U &=&SL(5),\quad K=SO\left( 5\right) =Sp(4) \\
T &=&SO(3,3),\quad k=SO\left( 3\right) _L\otimes SO\left( 3\right) _R=SO(4)
\nonumber \\
R^{\left( 2\right) } &=&\left( 6_{space}+3_{L,int}\right) \otimes \left(
6_{space}+3_{R,int}\right) ,  \nonumber
\end{eqnarray}
The $k=SO(3)\otimes SO(3)$ structure is obvious from the internal dimensions
$\left( 3_{int}\right) _{L,R}.$ Can these be put together into complete $%
K=SO\left( 5\right) $ multiplets? The answer is no, but this is not
surprizing since we have already seen that the $l=2$ perturbative states are
insufficient. Can we add sufficient number of non-perturbative states to
make complete $K$-multiplets? The answer is, of course, yes, but the needed
states go beyond the minimal extension (\ref{9to10}). There is a minimal
number of states that we can add to obtain complete U-multiplets, but these
do not agree with the minimal number that make $SO(10)$ multiplets.

These examples show that we can always find a minimal set of
non-perturbative states to make complete U-multiplets in accordance with eq.(%
\ref{fields}). However, the systematics of the missing states is not always
in accordance with the systematics of {\it minimal} number of 11D multiplets.

\section{Is there hidden 11D structure?}

In the previous section we saw that for
\[
\left( d,c\right) =\left( 10,0\right) ,\left( 9,1\right) ,\left( 8,2\right)
,\left( 6,4\right)
\]
U-duality is consistent with the presence of hidden 11-dimensional structure
at {\it all string levels}. However for the other values of
\[
(d,c)=\left( 7,3\right) ,\left( 5,5\right) ,\left( 4,6\right) ,\left(
3,7\right)
\]
we found that the index structure required by the conjectured $U$-duality is
different than the {\it minimal number} of 11-dimensional supersymmetry
multiplets at excited levels $l\geq 2$. Does this mean we have to give up
the idea that there is a hidden 11D structure? In view of recent arguments
in favor of 11D \cite{witten},\cite{townsend}-\cite{jhs2}, the idea of 11D
seems to become more compelling. If both U-duality and 11D are true then
together they must give powerful restrictions on the structure of the theory
from two different non-perturbative perspectives. The first test is to find
a resolution for the spectrum when the two requirements seem to conflict
with each other, as above. The only possible solution is that there exists a
collection of states that is bigger than the minimal set required by either $%
K$ or $SO(10)$. The property of this collection must be that it can be
reclassified either as $K$-multiplets or $SO(10)$ multiplets. When
classified as $SO(10)$ states the $K$ structure may be obscured or vice
versa. This is a testable hypothesis.

Thus, let's reconsider the case of $\left( d,c\right) =\left( 7,3\right) $
at level $l=2.$ The perturbative states are summarized by the 81 indices in
the following $SO\left( 6\right) \otimes \left[ SO\left( 3\right) _L\otimes
SO\left( 3\right) _L\right] $ representations (besides the 2$%
_B^{15}+2_F^{15} $ factor)
\begin{eqnarray}
R^{\left( 2\right) } &=&9_L\otimes 9_R\rightarrow \left(
6_{space}+3_{L,int}\right) \otimes \left( 6_{space}+3_{R,int}\right)
\nonumber \\
&=&\left( 1+15+20,\left[ \left( 0,0\right) \right] \right) + \\
&&\left( 6,\left[ \left( 3,0\right) +\left( 0,3\right) \right] \right)
+\left( 0,\left[ \left( 3,3\right) \right] \right)  \nonumber
\end{eqnarray}
We are seeking $SO\left( 6\right) \otimes SO(5)$ structures that are
compatible with a decomposition of $SO(10)$ multiplets. We have found 94
additional non-perturbative states that combine together with the 81
perturbative ones to give 175 states with the desired properties:
\begin{eqnarray}
SO(10) &:&1+3\times 10+2\times 45+54=175 \\
SO\left( 6\right) \otimes SO(5) &:&(0,0)+5\times \left( 6,0\right) +2\times
\left( 15,0\right) +\left( 20,0\right)  \nonumber \\
&&+\left( 6,10\right) +2\times \left( 0,10\right) +\left( 0,14\right)
\nonumber
\end{eqnarray}
To see that these match each other, and also include the 81 perturbative
states, we need to decompose them according to the following schemes
\begin{eqnarray}
&&^{SO(10)\rightarrow SO(6)\otimes SO(4)_1\rightarrow SO(6)\otimes SO(3)} \\
&&_{SO\left( 6\right) \otimes SO(5)\rightarrow SO(6)\otimes
SO(4)_2\rightarrow SO(6)\otimes SO(3)}  \nonumber
\end{eqnarray}
where $SO(4)_2=SO\left( 3\right) _L\otimes SO\left( 3\right) _L=k$ is the
subgroup of $T=O\left( 3,3\right) ,$ whereas the $SO(4)_1$ is different
since it does not contain the left/right information. However, the $SO(3)$
subgroup is common to both.

The minimal $SO(10)$ multiplets that came from $\left( 9+1\right) _L\otimes
\left( 9+1\right) _R$ are $1+45+54.$ These contain perturbative and
non-perturbative states. The additional ones $3\times 10+45$ are purely
non-perturbative. For the cases $\left( d,c\right) =\left( 10,0\right)
,\left( 9,1\right) ,\left( 8,2\right) ,\left( 6,4\right) $ the additional
ones were not required by the minimal completion. However, presumably these
additional multiplets are present also in those cases if there is a common
11D structure behind the whole theory.

In this example we have shown that it is possible for both U-multiplets and
11D multiplets to coexist even though they appeared to be at conflict at
first. It is quite difficult to carry out a similar analysis for other
levels $l$ and other cases $\left( d,c\right) .$ At this stage we have not
understood enough of the theory to know whether a similar conclusion is true
more generally. Therefore the issue remains open until a better approach is
found to answer the question: do U-duality and 11D structure coexist?
Hopefully the answer is yes, otherwise we have to decide which one is right.

\section{Summary and Comments}

In this work we have investigated the compactified type-IIA superstring for
various values of $\left( d,c\right) $. We have presented an algebraic
approach which encompasses both the perturbative as well as the
non-perturbative states by putting them into U-multiplets. We have reasoned
that the multiplet is labelled by two spaces, ``index'' space and ``base''
space. Both spaces are extensions of similar spaces that label the
perturbative T-duality multiplets. In our scheme U-transformations, much
like T-transformations, do not mix perturbative states from different levels
(note the definition of level for non-perturbative states, as given in the
introduction). Our whole analysis critically depends on this structure which
allows us to investigate the U-multiplets level by level. The paper
presented evidence that this scheme is consistent (to the extent of our
investigation).

We have found that at levels $l=0,1$ the existing index structure for
perturbative states is all that is needed to define complete $U$-multiplets
in the form $\Phi _{indices}^{\left( l\right) }(x^\mu ,base)$ for all values
of $\left( d,c\right) $, and that this result directly follows from the
simplest short and long multiplet structure of 11D space-time supersymmetry.

At levels $l=0,1$ all non-perturbative aspects appear in the $base=\left(
\vec{m},\vec{n},z^I\right) $. The base quantum numbers are the central
charges of the 11D SUSY algebra and these correpond to the 0-brane sources
that couple to the massless vector particles in supergravity. $U$ acts as a
linear transformation on the base in a representation that is identical to
the one applied to the massless vector fields in compactified 11D
supergravity. Furthermore the indices correspond to complete representations
of $K$ and they mix with a transformation induced by $U.$

Hence, for $l=0,1$ both index space and base space of U-multiplets have firm
connections to 11D.

To have $U$-duality at higher levels $l\geq 2$ additional non-perturbative
states are needed to complete the index structure. If these additional
states are absent in the theory there is no $U$-duality in the full theory.
We would then have to interpret the successful result at $l=0,1$ as a pure
accident. This is not impossible since the $l=0,1$ cases are fully explained
by SUSY. On the other hand, assuming that U-duality is true for $l\geq 2$,
our approach provides an algebraic tool for identifying the non-perturbative
states at every level once the perturbative states are listed.

There seems to be a non-perturbative 11D structure lurking behind the
theory. In view of the existence of a classical membrane theory with some
promise of its consistency at the quantum level, searching for hidden 11
dimensional structure is an interesting challenge. It is not necessary for
11D to be present in the 10D theory, but there is mounting evidence for it,
including the work we presented here. We have seen that U-duality is
distinct from this 11D structure, although in some cases they appeared to
imply each other. We have found cases where there is a clash between the two
if one is restricted to a minimal set of non-perturbative states. We have
shown at least in one example that the conflicts may be resolved by adding
more non-perturbative states. But nevertheless this example clearly shows
that 11D and U-duality are quite distinct from each other. If they are both
true their combined effect is quite restrictive on the non-perturbative
structure of the theory. Whether the conflict can be resolved generally is a
major question raised by our work.

Other places where our ideas could be tested is in the proposed dualities
between the Heterotic string on $R^6\otimes T^4$ and the type-IIA string on $%
R^6\otimes K_3,$ as well as other similar cases involving heterotic, type-I
or type-II theories. The perturbative plus non-perturbative spectrum of
these theories should match each other. By using the perturbative
T-multiplets of either theory as a starting point and then requiring
U-multiplets at each level one should find the same full spectrum from
either side.

In this paper we have not discussed multiplets with $p$-branes that also
enter the picture \cite{jhs1}-\cite{sen-u}. However, we propose to include
them algebraically as follows. Since $p$-branes are sources for $p+1$ forms
we can draw a parallel between the central charges for $p$-branes in the
SUSY algebra and the $p+1$ forms in compactified supergravity. By analogy to
the $p=0$ case which we have discussed, we expect that the $p$-brane charges
are classified in the same $K$ or $U$ multiplets corresponding to the $p+1$
forms. Carrying the analogy further we expect the base to include all the $p$%
-brane charges
\begin{equation}
base=\left( 0\text{-brane charges,\thinspace \thinspace }\cdots \,\,,\,\,p%
\text{-brane charges,}\cdots \right)
\end{equation}
Thus, we propose that the base consists of a U-multiplet for each $p$-brane.
It will be interesting to study $p$-branes and further explore this
possibility.

\section{Acknowledgements}

Both of us would like to acknowledge the hospitality and support of CERN
where this work was initiated. We would also like to thank M. Poratti for
discussion on T-duality.

\section{Appendix}

In this appendix we review the massless sector of the 10D type-IIA string
that coincides with the fields of 11D supergravity. We need to understand
the ``index'' structure of these fields as this will be the basis for the $U$%
-multiplet and $K$-multiplet structure at level $l=0.$ Since our aim is to
find the largest multiplet structure we will use 11-dimensional
classifications. The massless multiplets are the 11-dimensional graviton,
3-index antisymmetric tensor, and gravitino. In our notation these are the $%
l=0$ fields
\begin{equation}
\phi _{indices}^{\left( 0\right) }:g_{MN},\,\,A_{MNP},\,\,\psi _{M\vec{\alpha%
}}.\,
\end{equation}
In the lightcone gauge, the physical degrees of freedom of the string are
classified by SO(8) which is the little group for massless 10-dimensional
states. However, it is possible to regroup the SO(8) representations into
SO(9), which is the little group for massless states in 11-dimensions, by
taking $M,N=1,\cdots 9,$ while $\bar{\alpha}$ is the 16-dimensional SO(9)
spinor index. This well known fact is a first indication of a hidden extra
dimension.

If the string theory is toroidaly compactified to $R^d\times T^c,$ with $%
d+c=10,$ then the ``base'' acquires additional quantum numbers that
correspond to Kaluza-Klein, winding, and central charge quantum numbers,
while the ``indices'' must now be split into space and internal parts. In
the lightcone notation this corresponds to decomposing the SO(9)
representations above to $SO(d-2)\times SO(11-d)$ in order to obtain the
spin and internal symmetry content of the fields. The indices split as
follows
\begin{eqnarray}
M &\rightarrow &m\oplus i\quad \bar{\alpha}\rightarrow \alpha a  \nonumber \\
m &=&1,2,\cdots ,d-2\quad \leftrightarrow SO(d-2)  \nonumber \\
i &=&1,2,\cdots ,c+1\quad \leftrightarrow SO(c+1)=SO(11-d) \\
\alpha &=&\text{Spinor of }SO(d-2)  \nonumber \\
a &=&\text{Spinor of }SO(11-d)  \nonumber
\end{eqnarray}
Note that $c$ has been augmented by ``1'' due to the hidden compactified
11th dimension. Then the fields are decomposed as follows
\begin{eqnarray}
g_{MN} &\rightarrow &g_{mn}\oplus V_m^i\oplus S^{\left( ij\right) }
\nonumber \\
A_{MNP} &\rightarrow &A_{mnp}\oplus B_{mn}^i\oplus V_m^{\left[ ij\right]
}\oplus S^{\left[ ijk\right] }  \label{decomp} \\
\psi _{M\bar{\alpha}} &\rightarrow &\psi _{m\alpha }^a\oplus \psi _\alpha
^{ia}  \nonumber
\end{eqnarray}
where we have written the space indices as subscrips and internal indices as
superscripts. From the point of view of spins there are scalars, vectors,
tensors, and a variety of spinors. We can classify them according to $%
SO\left( d-2\right) \times SO\left( c+1\right) $ since this can be read-off
directly from the indices above. We can count them, and obtain their total
numbers to see in which representations of $U$ or $K$ they would fit, as
explained in table II. The counting has to take into account that in some
dimensions a tensor may be dual to a vector or scalar, etc. After this is
taken into account we obtain the classsifications given in the tables below
for various values of $\left( d,c\right) .$

\begin{eqnarray*}
&&
\begin{tabular}{|l|l|}
\hline
$\frac dc\rightarrow \frac{d-2}{c+1}$ & $
\begin{array}{c}
SO(d-2)\text{ \thinspace SCALARS} \\
S^{\left( ij\right) }\oplus S^{\left[ ijk\right] }\oplus S^{duals}
\end{array}
$ \\ \hline
9/1$\rightarrow $7/2 & $\frac{2.3}2+0+0=3$ \\ \hline
8/2$\rightarrow $6/3 & $\frac{3.4}2+\frac{3.2.1}{1.2.3}+0=7$ \\ \hline
7/3$\rightarrow $5/4 & $\frac{4.5}2+\frac{4.3.2}{1.2.3}+0=14$ \\ \hline
6/4$\rightarrow $4/5 & $\frac{5.6}2+\frac{5.4.3}{1.2.3}+0=25$ \\ \hline
5/5$\rightarrow $3/6 & $\frac{6.7}2+\frac{6.5.4}{1.2.3}+1\left( \tilde{A}%
_{mnp}\right) =42$ \\ \hline
4/6$\rightarrow $2/7 & $\frac{7.8}2+\frac{7.6.5}{1.2.3}+7\left( \tilde{B}%
_{mn}^i\right) =70$ \\ \hline
3/7$\rightarrow $1/8 & $\frac{8.9}2+\frac{8.7.6}{1.2.3}+\left( 8+28\right)
\left( \tilde{V}_m^i+V_m^{\left[ ij\right] }\right) =128$ \\ \hline
\end{tabular}
\\
&&\text{TABLE V. The number of scalars is dim}\left( U/K\right) .
\end{eqnarray*}

\begin{eqnarray*}
&&
\begin{tabular}{|l|l|}
\hline
$\frac dc\rightarrow \frac{d-2}{c+1}$ & $
\begin{array}{c}
SO(d-2)\,\,\,\text{VECTORS} \\
V_m^i+V_m^{\left[ ij\right] }+V_m^{duals}
\end{array}
$ \\ \hline
9/1$\rightarrow $7/2 & $2+1+0=3$ \\ \hline
8/2$\rightarrow $6/3 & $3+\frac{3.2}2+0=6\rightarrow \left( 3,2\right) $ \\
\hline
7/3$\rightarrow $5/4 & $4+\frac{4.3}2+0=10$ \\ \hline
6/4$\rightarrow $4/5 & $5+\frac{5.4}2+1\,\,(\tilde{A}_{mnp})=16$ \\ \hline
5/5$\rightarrow $3/6 & $6+\frac{6.5}2+6\,\,\left( \tilde{B}_{mn}^i\right)
=27 $ \\ \hline
4/6$\rightarrow $2/7 & $7+\frac{7.6}2+0=28=56\,\,($self dual) \\ \hline
3/7$\rightarrow $1/8 & $0,\,\,($dual to scalars $V^i+V^{ij})$ \\ \hline
\end{tabular}
\\
&&\text{TABLE VI. Dimensions of }U_{global}\text{ multiplets.}
\end{eqnarray*}

\begin{eqnarray*}
&&
\begin{tabular}{|l|l|}
\hline
$\frac dc\rightarrow \frac{d-2}{c+1}$ & $\left.
\begin{array}{c}
SO(d-2)\,\,\,\text{TENSORS} \\
B_{mn}^i+B_{mn}^{duals}
\end{array}
\right. $ \\ \hline
9/1$\rightarrow $7/2 & $2+0=2$ \\ \hline
8/2$\rightarrow $6/3 & $3+0=3$ \\ \hline
7/3$\rightarrow $5/4 & $4+1$ ($\tilde{A}_{mnp})=5$ \\ \hline
6/4$\rightarrow $4/5 & $5+0=5=10\,\,$self dual \\ \hline
5/5$\rightarrow $3/6 & 0, dual to vector $B_m^i$ \\ \hline
4/6$\rightarrow $2/7 & 0, dual to scalar $B^i$ \\ \hline
3/7$\rightarrow $1/8 & 0 \\ \hline
\end{tabular}
\\
&&\text{TABLE VII. Dimensions of }U\text{ multiplets.}
\end{eqnarray*}

\begin{eqnarray*}
&&
\begin{tabular}{|l|l|}
\hline
$\frac dc\rightarrow \frac{d-2}{c+1}$ & $\left.
\begin{array}{c}
\text{GRAVITINOS\thinspace }\,\psi _{m\alpha }^a\,\,\text{under} \\
SO(d-2)\otimes K
\end{array}
\right. $ \\ \hline
10/0$\rightarrow $8/1 & $56_{+}+56_{-}$ \\ \hline
9/1$\rightarrow $7/2 & $\left( 48,\pm \right) $ \\ \hline
8/2$\rightarrow $6/3 & $\left( 20,2,+\right) +\left( 20^{*},2,-\right) $ \\
\hline
7/3$\rightarrow $5/4 & $\left( 16,4\right) $ \\ \hline
6/4$\rightarrow $4/5 & $\left( \left( 2,3\right) ,\left( 4,0\right) \right)
+\left( \left( 3,2\right) ,\left( 0,4\right) \right) $ \\ \hline
5/5$\rightarrow $3/6 & $\left( 4,8\right) $ \\ \hline
4/6$\rightarrow $2/7 & $\left( -,8\right) +\left( +,8^{*}\right) $ \\ \hline
3/7$\rightarrow $1/8 & $0,$ dual to fermion $\psi _\alpha ^a$ \\ \hline
\end{tabular}
\\
&&\text{TABLE VIII. Dimensions of }K\text{ multiplets.}
\end{eqnarray*}

\begin{eqnarray*}
&&
\begin{tabular}{|l|l|}
\hline
$\frac dc\rightarrow \frac{d-2}{c+1}$ & $\left.
\begin{array}{c}
\text{FERMIONS}\,\,\psi _\alpha ^{ia},\text{ under} \\
SO(d-2)\otimes K
\end{array}
\right. $ \\ \hline
10/0$\rightarrow $8/1 & $8_{+}+8_{-}$ \\ \hline
9/1$\rightarrow $7/2 & $\left( 8,\pm \right) +\left( 8,\pm \right) $ \\
\hline
8/2$\rightarrow $6/3 & $\left( 4,4,+\right) +\left( 4^{*},4,-\right) +\left(
4,2,+\right) +\left( 4,2,-\right) $ \\ \hline
7/3$\rightarrow $5/4 & $\left( 4,16\right) $ \\ \hline
6/4$\rightarrow $4/5 & $\left.
\begin{array}{c}
\left( \left( 2,0\right) ,\left( 16,0\right) \right) +\left( \left(
2,0\right) ,\left( 4,0\right) \right)  \\
+\left( \left( 0,2\right) ,\left( 0,16\right) \right) +\left( \left(
0,2\right) ,\left( 0,4\right) \right)
\end{array}
\right. $ \\ \hline
5/5$\rightarrow $3/6 & $\left( 2,48\right) $ \\ \hline
4/6$\rightarrow $2/7 & $\left( +,56\right) \oplus \left( -,56^{*}\right) $
\\ \hline
3/7$\rightarrow $1/8 & $128\,\,\,\,\left( \psi _\alpha ^{ia}+\psi _\alpha
^a\right) $ \\ \hline
\end{tabular}
\\
&&\text{TABLE IX. Dimension of }K\text{ multiplets.}
\end{eqnarray*}

\end{document}